Effect of Night Laboratories on Learning Objectives for a Non-Major Astronomy Class

Ian C. Jacobi[1], Heidi Jo Newberg[1], Darren Broder[2], Rose A. Finn[2], Anthony J. Milano[1], Lee A. Newberg[1,3], Allan T. Weatherwax[2], and Douglas C. B. Whittet[1]

[1]Rensselaer Polytechnic Institute, Troy NY USA; [2]Siena College Loudonville NY USA; [3]Wadsworth Center, New York State Department of Health, Albany NY USA

Abstract

We tested the effectiveness on learning of hands-on, night-time laboratories that challenged student misconceptions in a non-major introductory astronomy class at Rensselaer Polytechnic Institute. We present a new assessment examination used to assess learning in this study. We were able to increase learning, at the 8.0 sigma level, on one of the moon phase objectives that was addressed in a cloudy night activity. There is weak evidence of some improvement on a broader range of learning objectives. We show evidence that the overall achievement levels of the four sections of the class is correlated with the amount of clear whether the sections had for observing, even though the learning objectives were addressed primarily in activities that did not require clear skies. This last result should be confirmed with future studies. We describe our first attempt to cycle the students through different activity stations in an attempt to handle 18 students at a time in the laboratories, and lessons learned from this.

1. INTRODUCTION

       We describe the results of an experiment to see whether hands-on, nighttime astronomy laboratories improve student learning in a non-majors observational astronomy class at Rensselaer Polytechnic Institute. Our challenge was to design a meaningful laboratory experience, for a relatively large astronomy class of non-majors, that could be used in a site that suffers from poor weather, dome seeing, and significant light pollution. With few nights available for observation, we needed to be able to acquire targets quickly and store results digitally for later analysis. For our more advanced majors, some of whom go on to become successful professional astronomers, we wanted a professional control system and remote access to the observatory, so that it could be used in cold weather.

       We know of no other study that has been done on the effects of a nighttime activity-based learning pedagogy; most studies into the effects of activity-based learning are integrated as a component of the lecture time in the form of in-class activities or lecture-tutorials (for example McCrady and Rice, 2008). Both Straits & Wilke (2003), and Adams & Slater (1998) found that students generally like activities-based learning and Straits & Wilke (2003) saw the most improvement with activities that aided visualization of motion and spatial relations. No studies have focused on the effects of

activities that include use of a professional observatory setup on learning quality. In fact, Slater (2008) states that although people want to look through telescopes, "It is generally accepted that most undergraduate non–science majors taking an astronomy course spend little to no time actually looking through a telescope." The current focus is more in the direction of internet-accessible robotic telescopes (see Gould, Dussault, and Sadler, 2007, and references therein). In this paper we assess whether students' learning can is improved by moving the activities from the lecture or computer terminal to the real sky.

2. PREPARATIONS

2.1. Assessment Design

Our primary objective was to determine whether the establishment of a laboratory component to the introductory astronomy course (implemented independently but cooperatively between Siena and Rensselaer) would improve student achievement. To this end we created an assessment test in the summer of 2006, which was specifically aligned with the concepts taught in the introductory astronomy course at Siena College and the *Earth and Sky* class at Rensselaer. Since the assessment test was written a year before the night laboratories were written, it was not designed specifically for the laboratories, but rather on general learning objectives of the classes. Although the Astronomy Diagnostic Test (ADT, Zeilik, 2003) was available for our use, we felt that it did not align well enough to the course content to be a reliable indicator of student performance.

The results of the related education and research assessment at Siena College, which used the same assessment test, will be published separately. It should be noted that during the 2007–2008 academic year, both the assessment examination and the ADT were administered to students in the introductory astronomy course at Siena College. Since the ADT is largely regarded as the standard diagnostic instrument for introductory astronomy courses, we are interested to compare student performance on the assessment exam to student performance on the ADT.

Our assessment test was reviewed by three Siena faculty members and two Rensselaer faculty members in several iterations, including face-to-face meetings. In these meetings we attempted to ensure the clarity of each question and the relevance to the classes at Siena and Rensselaer. The resulting assessment test is presented in Appendix A.

The assessment test was administered to all students, twice, as a pretest and as a posttest, in the introductory astronomy *Earth and Sky* course at Rensselaer Polytechnic Institute during the Fall 2006 semester, prior to the creation and implementation of a formal laboratory component. It should be noted that the 16" telescope was available for use by the astronomy class during this semester, but it was used without the to-be-developed laboratories, and with partially compromised optics. While the pretest did not count towards the calculation of the students' score in the class, the posttest was counted as a portion of the final grade.

In this paper we compare the data collected from Fall 2006 with data collected after the inclusion of the laboratory component, in Fall 2007, in order to determine whether student achievement was enhanced.  Nighttime laboratory exercises were created for use with the facilities at Rensselaer's Hirsch Observatory that addressed some, but not all, of the questions in the assessment test.  The remaining questions on the assessment test were addressed in the lecture portion of the *Earth and Sky* class; the lecture portion was not changed between Fall 2006 and Fall 2007.

2.2. Telescope Equipment and Refurbishment

The main telescope in Rensselaer's Hirsch Observatory is a 16" B&C telescope that was built in 1965.  Until recently, the telescope had all of its original electromechanical components, which had become unreliable and difficult to use.  We purchased a CCD camera for the telescope, but were unable to use it in a practical way because it was difficult to point precisely enough at each object and because intermittent shorts in the electronics made it difficult to stay on a target once it was acquired. The observatory also owns several smaller telescopes that can be moved onto an adjacent roof area; these are often used for classes or public observing since only six people can fit in the dome of the 16" telescope at a time.

We decided to refurbish our existing telescope and dome so that it could be controlled both from within the dome area and from an adjacent warm room.  The refurbishment was done by Peter Mack from Astronomical Consultants and Equipment, Inc. (ACE).  It included replacing some of the drive motors; installing encoders for pointing and focus; and integrated, computerized control of the pointing, dome, and focus functions.  We are now able to acquire targets very quickly, we are able to track long enough to take very nice CCD images, and the system is regularly operated by our graduate and undergraduate students.  We have had a few mechanical failures, primarily related to the opening and closing of the dome, but these have been fixed and the system has been in regular use for the past year and a half, even through the mechanical difficulties.

The observatory is used by undergraduate classes, the Rensselaer Astrophysical Society undergraduate club, and the public observing program, which is staffed by a local amateur astronomer and a large group of volunteers that include Rensselaer students, Rensselaer staff, and interested local amateur astronomers of all ages.  We are in the process of making our campus observatory compliant with the Americans with Disabilities Act (ADA) by creating experiences for mobility impaired students and visitors to participate in the public and class activities from a wheelchair-accessible classroom which has a video and intercom link to the 16" telescope dome and which allows people to view the CCD images collected in real time from the telescope.  This would not have been possible before the upgrade of the telescope.  Students in classes will be able to use this facility and also control the telescope from a laptop located in the same room.

| Activity Number | Title | Topic |
|---|---|---|
| 1a | CCD Observing | Introducing the student to the telescope and CCD camera. |
| 1b | Field of View of a Small Telescope | Determining the field of view of a small telescope |
| 1c | Finding Stars and Constellations | Learning how to locate stars and constellations in the sky |
| 2a | Color CCD Images | Introducing the student to methods of creating pseudo-color CCD images |
| 2b | Aperture Size and Resolution | Determining the effects of aperture size on telescope observations |
| 2c | Atmospheric Extinction and Light Pollution | Illustrating the effects of the atmosphere on astronomical observations |
| 3a | Imaging a Cluster | Allowing the students to image an open cluster for a later activity. Learning about clusters. |
| 3b | Finding Planets | Finding the locations of planets in the night sky |
| 3c | Star Colors | Discovering the differences between the visible colors of stars |
| 4a | Imaging a Galaxy | Allowing the students to image a galaxy and study its color properties |
| 4b | Sky Scavenger Hunt | Finding the most astronomical objects in a limited amount of time. |
| 4c | Artificial Satellites | Discovering how satellites move in the sky and the reasons they are visible on Earth. |
| 5 | Light and Shadow in the Solar System | Visualizing how the Moon and Earth move in their orbits and how their orbits can cause moon phases and eclipses |
| 6 | Planning Observations | Discovering the calculations that a professional astronomer must make to have a successful observation. |
| 7 | Building a Scale Model of the Solar System | Gaining an understanding of the scale of the Solar System and universe |
| 8 | Making an H-R Diagram | Understanding the relationship between color and luminosity in H-R diagrams. |
| 9 | An Introduction to the CCD | Better understanding principles behind how a CCD works. |

*Table 1. List of implemented laboratory activities*

2.3. Laboratory Design

In addition to the refurbishment of the main telescope in the observatory, we needed a grander plan to bring meaningful laboratory experiences to 70-80 students in our non-majors class. We divided the class into four sections of about 18 students each. Each section met one night every other week, so this class could be accommodated in two nights per week. Each section was then subdivided into three groups of six students, and these groups rotated through three activities in each clear night. The stations for the three simultaneous activities were (1) the 16" telescope, (2) the small telescopes on the adjacent rooftop area, and (3) a self-guided naked eye activity. If the assigned night was cloudy, we provided a separate set of labs that could be done either in the observatory or in a classroom.

We designed four clear night activities, each of which has three parts (a, b, c), and five activities that could be done in cloudy weather. We expected that this would be enough activities to cover a diversity of weather conditions during the laboratory times. These activities can be found at http://www.rpi.edu/dept/phys/observatory/labs.html, and the titles of the activities are listed in Table 1.

One thing we learned in making the activities is that it is difficult to design laboratories that challenge misconceptions or reinforce astronomical concepts that are typically taught in non-major courses using a telescope, and particularly an automated telescope with a CCD camera. Although students really like using the big, automated telescope, they are not challenged to confront their misconceptions by using it. Using the smaller telescopes, they are more likely to discover that the sky appears to move, or that the size, alignment and focus of the optics matter, or that the North Star stays in the same place. All of this is handled for them on the automated telescope. It was much easier to teach standard concepts about the phases of the moon or the paths of the planets or the effects of the atmosphere in naked eye and cloudy night laboratories. The important advantage of the 16" telescope was that it was a big draw that the students all wanted to use.

The primary objective of many of the 16" laboratories was the development of astronomical skills through hands-on interactions, but the laboratories were designed such that questions on the assessment were also addressed simultaneously. Other laboratories, especially those that utilized naked eye observations, focused specifically on reducing common astronomical misconceptions, especially those covered by the assessment, by directly questioning the students before, during, and after the laboratory was performed through the use of pretests and posttests.

3. METHODOLOGY

3.1. Participants and Setting

This education research project was staffed in an unusual way. The coordination of the refurbishment of the telescope and the creation and testing of the astronomy laboratories were done primarily by a professor (HJN) and an undergraduate student (ICJ) who never taught the class and were not available during the laboratories. The lecture, homework, and testing part of the class were run by a different professor (DCBW), who has taught that part of the class the same way for a number of years. The night laboratories were staffed by undergraduate students and first year graduate students.

Rensselaer's *Earth and Sky* class enrolls 70-80 students and is taught during the fall semester only. Before this study, the class met in a standard lecture format with homework and exams. Traditionally there had been a laboratory component that consisted of attending a minimum of five clear nights at our campus observatory. Before the implementation of the night laboratories, all of the telescope-related activities were designed and run by a single graduate student. At these observation nights the students would view three objects per night, draw what they observed in the telescope, and provide a description of each object based on the explanation given by the graduate student running the observing sessions. The students generally enjoyed these activities, but many of the students had difficulty planning ahead far enough to attend five clear observing nights.

The new night laboratories operated two nights per week, and on each night a team of one graduate student and one undergraduate student ran the activities. Typically, the graduate student was responsible for running the activity on the 16" telescope, the undergraduate student was responsible for running the small-telescope activity, and the naked-eye activities were self-directed. None of the staff that taught the night laboratories had significant previous teaching experience, but all four had had some previous exposure to astronomy.

Even with the significant extra staffing over the previous year, the new laboratories were difficult to implement. It is not easy to convince undergraduate students that they should confront misconceptions and think through the lessons in a group activity. Allen and Kelly-Riley (2006) note this problem when they attempt to get students to think critically about their laboratory exercises rather than doing them mechanically without thinking. Although most of them enjoy looking at objects through the telescope, Rensselaer students see each assignment as an obstacle to be overcome as quickly as they can, mechanically and without thinking if they can get away with it. The graduate and undergraduate teaching assistants were not always prepared to keep them focused on learning the material. Although the plan was to have two small telescopes on the roof so that three students would be cooperating with each other on each of them, with an undergraduate TA to help set them up, the students quickly put pressure on the undergraduate TA to set up one telescope and walk them all through the activity together, so that they didn't have to do as much tedious work learning how to use the telescope. Students assigned to the naked-eye activity, who were using the same rooftop, often tagged along with the small telescope group so that there were eight or ten students hanging around one small telescope and learning the answers from one another. Running the laboratories this way, the small-telescope and naked-eye groups always finished ahead of the group in the 16" telescope dome, and then had to wait. There was significant pressure on the TAs to get the students out of the lab in well under 3 hours, especially since the students were spending some of that time just waiting.

The teaching assistants also did not fully grasp the idea of groups cycling through stations. They tended to use the prepared activities as a set of resources from which they then improvised a lesson plan. For instance, they would discover that it was possible to answer the questions in many of the naked-eye laboratories without actually looking at the sky; one activity asked the students to use a planisphere and compare it with the sky, but most of the questions could be answered even if the sky was not there to compare with. So the students sometimes plowed through a set of naked-eye laboratories on a cloudy night. Since each of the four laboratory sections was doing a different lab (depending on which groups did or did not have good weather) and since sometimes the TAs were mixing parts of one lab with another lab, the TAs had some difficulty keeping track of who had done what and when.

3.2. Data Collection

*3.2.1. Assessments*

At the beginning and end of each of the Fall 2006 and Fall 2007 semesters, the assessment tests were given to the students in the *Earth and Sky* course in order to measure the improvement in the quality of astronomical knowledge held by students through the course. In particular, pretests were given soon after the start of the course and were not included in the final grade, and posttests were given as part of the final for all courses and integrated into the final grade.

*3.2.2. Solicited Feedback and Course Evaluations*

Over the course of the Fall 2007 semester, feedback on the design and opinions of the laboratories were solicited from students in the Rensselaer classes. Feedback was solicited both in the *Earth & Sky* class as well as by e-mail. In addition, the impact of the nighttime laboratories was assessed through comparisons of the responses from the standard Individual Development & Education Assessment (IDEA) Diagnostic Forms (Hoyt & Lee, 2002, http://www.idea.ksu.edu/) that are regularly distributed in all Rensselaer classes.

4. ANALYSIS AND FINDINGS

4.1 Assessment Test

As not all of the students in the class remained in it for the entirety of the semester, we first chose to remove from the sample those tests given to individuals who did not take both a pretest and a posttest. This resulted in a cleaner data set that more clearly illustrates the change in the performance of the sample between the pretest and posttest, as it is properly normalized against the performance of the same individuals in both groups. There were 82 students who took both the pretest and the posttest in Fall 2006, and there were 68 students who took both in Fall 2007.

We then examined the similarities between the pretest samples given in 2006 and 2007 in order to ascertain whether the two sample groups could be considered part of the same population, which is an issue when one makes significant changes to an elective course. If the scores are consistent with being drawn from the same student population then the pretest sample can be taken to be the union of the 2006 and 2007 samples, thus reducing the error in the pretest statistics. If the students in the two classes are not similar, then our ability to measure learning improvement is somewhat compromised, since we are then comparing the effect of new a new teaching technique in classes that are different in the ability or preparation of the students. One can attempt to correct for differences between classes, but it cannot be done as cleanly; for example Brogt et al. (2007) show that for some measures of learning gain, there can be a bias in the measured learning for different pretest scores.

Table 2 lists the number of correct answers to each pretest question in 2006 and 2007. The fraction ($p$) of the students who chose the correct answer is found by dividing the number of correct answers in 2006 by 82 students, and by dividing the number of correct answers in 2007 by 68 students. The underlying probability of selecting the correct answer is estimated as the fraction of the students who selected it, and the assumed probability distribution is given by the binomial distribution function:

$$f(k;n,p) = \frac{n!}{k!(n-k)!}p^k(1-p)^{n-k},$$

where $n$ is the number of students having taken the particular assessment test for the given year, $p$ is the probability of selecting the answer, the mean number of correct answers is $np$, and the variance is $np(1-p)$. If we are using the tests to determine $p$, the probability of selecting the correct answer, then the error in $p$ is $\sigma$=sqrt[$p(1-p)/(n-1)$].

We used this last formula to determine the number of standard deviations between the probabilities of answering each pretest question correctly:

$$z_{TOT} = (p_{2007}-p_{2006})/\text{sqrt}(\sigma_{2006}^2 + \sigma_{2007}^2),$$

evaluating whether these values would be distributed like a Gaussian centered at zero with a standard deviation of unity. Note in the present calculation we are comparing pretests only; the $p$'s and $\sigma$'s are for the pretests. The distribution of the $z_{TOT}$ values over the 25 questions had a mean of 0.35±0.26 and a standard deviation of 1.31. Both the standard deviation and the mean were slightly higher than expected, but at much less than the two-sigma level. The 2007 class did slightly better on the pretest overall, but more than four sigma worse on question 23. Within the errors, it was difficult to draw a firm conclusion as to whether the two classes' pretest assessments were drawn from the same population.

We examined the entire distribution of responses, including the distribution of responses on the incorrect answers, and concluded that there was not an obvious difference between the response patterns. Combining this information with the very weak evidence for a population difference between the classes we decided to assume the populations were the same.

| Q# | Number Correct in 2006 Pretest | Number Correct in 2006 Posttest | Number Correct in 2007 Pretest | Number Correct in 2007 Posttest | Fractional Change in 2006 | Fractional Change in 2007 | Std. Dev. of 2006 Change | Std. Dev. of 2007 Change | Z |
|---|---|---|---|---|---|---|---|---|---|
| 1 | 49 | 57 | 44 | 49 | 7.5% | 10.1% | 6.5% | 6.8% | 0.3 |
| 2 | 20 | 33 | 16 | 19 | 16.2% | 3.9% | 6.5% | 6.5% | -1.6 |
| 3 | 26 | 41 | 23 | 43 | 17.3% | 30.6% | 6.8% | 7.0% | 1.6 |
| 4 | 38 | 19 | 37 | 22 | -26.8% | -17.6% | 6.2% | 7.0% | 1.2 |
| 5 | 55 | 59 | 53 | 53 | 0.0% | 5.9% | 6.2% | 6.3% | 0.8 |
| 6 | 48 | 67 | 52 | 56 | 15.0% | 15.7% | 5.8% | 6.1% | 0.1 |
| 7 | 51 | 63 | 42 | 58 | 14.8% | 23.3% | 6.1% | 5.9% | 1.3 |
| 8 | 19 | 43 | 22 | 39 | 25.1% | 30.0% | 6.6% | 7.1% | 0.6 |
| 9 | 8 | 24 | 7 | 21 | 19.3% | 20.9% | 5.6% | 6.2% | 0.2 |
| 10 | 53 | 63 | 48 | 58 | 9.5% | 18.0% | 6.1% | 5.8% | 1.3 |
| 11 | 32 | 23 | 35 | 56 | -16.6% | 37.7% | 6.4% | 6.2% | 8.0 |
| 12 | 43 | 53 | 41 | 47 | 8.6% | 13.1% | 6.7% | 7.0% | 0.6 |
| 13 | 59 | 69 | 62 | 61 | 3.5% | 9.0% | 5.2% | 4.9% | 1.0 |
| 14 | 68 | 74 | 61 | 63 | 4.2% | 6.6% | 4.4% | 4.3% | 0.5 |
| 15 | 20 | 30 | 19 | 26 | 10.6% | 12.2% | 6.4% | 6.9% | 0.2 |
| 16 | 35 | 30 | 20 | 37 | -0.1% | 17.7% | 6.7% | 7.3% | 2.2 |
| 17 | 65 | 74 | 58 | 66 | 8.2% | 15.1% | 4.6% | 3.8% | 1.8 |
| 18 | 34 | 49 | 24 | 41 | 21.1% | 21.6% | 6.8% | 7.2% | 0.1 |
| 19 | 56 | 67 | 49 | 55 | 11.7% | 10.9% | 5.7% | 6.1% | -0.1 |
| 20 | 77 | 77 | 63 | 62 | 0.6% | -2.2% | 3.4% | 4.0% | -0.6 |
| 21 | 42 | 44 | 41 | 46 | -1.7% | 12.3% | 6.9% | 7.0% | 1.8 |
| 22 | 41 | 52 | 36 | 43 | 12.1% | 11.9% | 6.7% | 7.2% | 0.0 |
| 23 | 32 | 26 | 9 | 12 | 4.4% | -9.7% | 6.3% | 5.9% | -2.0 |
| 24 | 30 | 29 | 22 | 26 | 0.7% | 3.6% | 6.6% | 7.1% | 0.4 |
| 25 | 28 | 52 | 24 | 38 | 28.7% | 21.2% | 6.6% | 7.2% | -0.9 |
| Direct | 318 | 372 | 285 | 371 | 6.5% | 17.9% | 2.4% | 2.5% | 4.1 |
| " w/o 11 | 286 | 349 | 250 | 315 | 9.8% | 15.1% | 2.6% | 2.7% | 1.8 |
| Indirect | 292 | 325 | 266 | 294 | 3.5% | 8.6% | 2.6% | 2.7% | 1.7 |
| Both w/o 11 | 578 | 674 | 516 | 609 | 6.6% | 11.9% | 1.8% | 1.9% | 2.5 |
| Unaddressed | 419 | 521 | 357 | 432 | 11.8% | 11.8% | 2.1% | 2.3% | 0.0 |

*Table 2. Fractional Change in Probability of Obtaining Correct Answers. Fractional change is calculated using combined pretest sample, rather than that for individual year.*

The baseline probability ($p_{PRE}$) of obtaining the correct answer for each question was calculated by adding the number of correct pretest responses in 2006 and 2007 and dividing by 150 (the total number of students taking the pretest). The probability of obtaining the correct answer to each question on the posttest was calculated separately for each year, dividing the number of correct answers (see Table 2) to each question by 82 for 2006, and by 68 for 2007. The fractional change is ($p_{POST}$-$p_{PRE}$)×100%. Columns 8 and 9 in Table 2 give the standard deviation of the fractional change for 2006 and 2007, computed by adding the standard deviations of pretest and posttest probabilities in quadrature. Note that the fractional change for 2006 and 2007 have correlated errors, since they both use the same pretest scores. When we compare the fractional changes by taking their difference, we are effectively only comparing the posttest results, since we used the same (averaged) pretest results. Therefore, the standard deviations of the pretests were not used when computing the numbers in the last column, which give the number of standard deviations by which the 2007 results differ from the 2006 results. $Z$ is computed by dividing the difference between the 2007 and 2006 fractional changes by the standard deviations of the probabilities of selecting the correct answer (on the posttest only from 2006 and 2007) added in quadrature. For example, on Question 1, there is a 0.3 sigma difference between the results between 2006 and 2007, which is not a significant improvement.

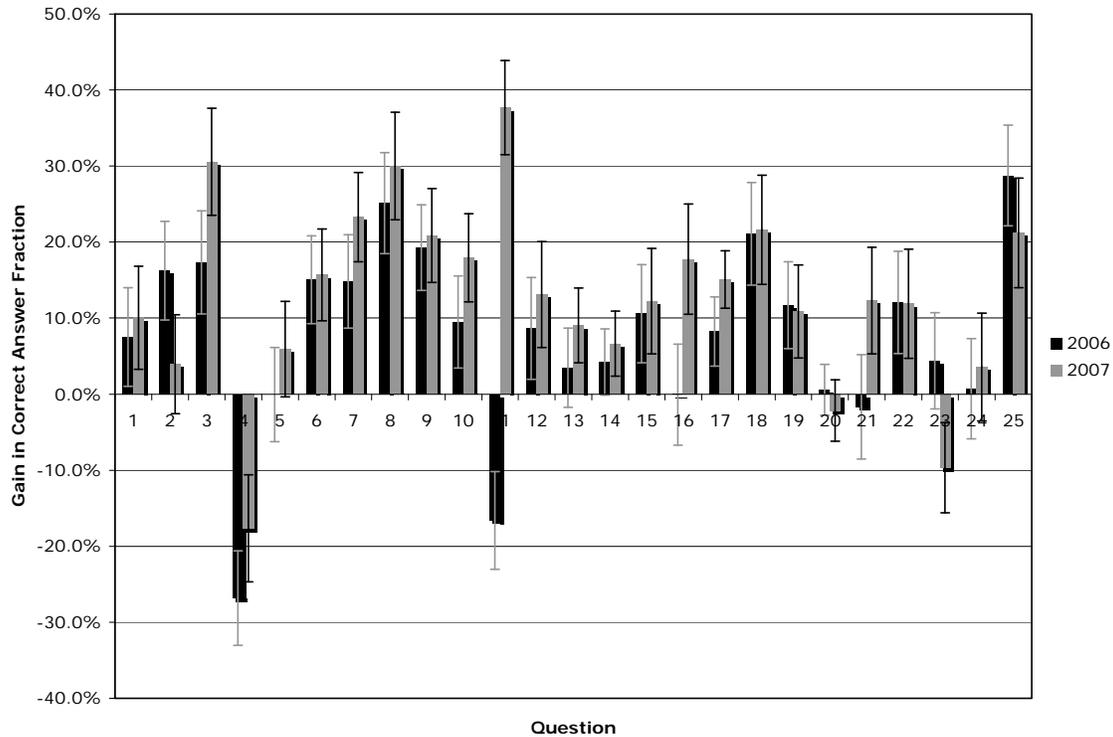

*Figure 1. Change in the probability of obtaining the correct answer between the averaged 2006-2007 pretest sample and the posttest sample of a given year, by question number.*

Figure 1 shows the fractional change in 2006 and 2007 broken down by question. The most notable results in Figure 1 are the remarkable improvement in question 11 in 2007 after the laboratories were introduced, as well as the decrease in both years of obtaining the correct answer for Question 4. The learning on Question 11 differed between the two years at the eight sigma level. No other question has a difference of more than 2.5 sigma. In addition to the one question which was tremendously affected by the learning labs, we find that the fractional change is greater in 2007 in 19 of the remaining 24 questions, suggesting that there may be a smaller effect on a each of number of the learning goals. We will address the combined results from multiple questions later in this section.

Question 4's decrease in correct answer probability is unusual not only for being one of few cases in which there was actually a decline in the probability of obtaining the correct answer following the course, but also for being the only case where this decline was present both before and after the night laboratories were included. Figure 2 shows the frequency of each possible answer to this question (a – e, plus one student who left the answer blank) in each of the four times the exam was given. The possible responses are written in Appendix A. This question tests whether students know that the resolving power of the telescope depends on both the size of the telescope aperture and the wavelength (color) of the light being observed. After taking the course, more students knew that the size of the telescope aperture affected the resolving power (answer `a'), but fewer students recognized that there is also a dependence on color (answer `c'). It is not known whether the problem is that students did not relate color to wavelength (the wavelength change is a small effect in the visible region of the spectrum), or whether they did not know that wavelength was a concern. Perhaps they were confused about whether the color referred to the color of the star. The shifting of support to answer 'a' rather than the correct answer, `c,' may indicate that the course itself stressed the importance of the effect of the size of the aperture on telescope resolving power, but not the effect of the wavelength of the light being observed, which serves as the other part of the correct answer. We will consider re-writing this question on the assessment test.

We now turn to the question of whether there was a significant improvement in the learning objectives, other than in Question 11, due to the introduction of the night laboratories. Given that the laboratory assignments did not explicitly focus on every question, we chose to examine the change in probability by grouping the questions by their status as having been directly addressed (questions 1, 3, 6, 9, 11, 12, 16, and 20), indirectly addressed (4, 7, 10, 13, 15, 22, and 24), or not addressed at all by any lab (2, 5, 8, 14, 17, 18, 19, 21, 23, and 25). Question 21 was classified as "Not Addressed," because it was addressed only in Lab 4c, which was not completed by any of the sections. Lab 4c was not developed in time to be given to the class in 2007, but is available for future years. Due to the extreme outlier-effect that Question 11 was likely to introduce into the findings, we also grouped all directly addressed questions together excepting Question 11. The questions, the labs that addressed them, and the level at which they were addressed are given in Table 3. The change in the probability of obtaining a correct answer for a question in this grouping is given in Figure 3, and tabulated at the end of Table 2.

From the average change in the probability of obtaining a correct answer as categorized by question type, we are able to conclude that there was a statistically significant improvement in all questions addressed by the lab activities, with a 4.1 sigma difference between the improvement in 2006 and the improvement in 2007. If we exclude Question 11, the significance of the improvement in directly addressed learning objectives drops to 1.8 sigma, which is pretty marginal. (For this calculation, we treated Question 11 as if it had never existed. The result would have been slightly stronger if we had tried to evaluate a null statistical model under which the strongest question among 25 questions is always excluded.) The significance of the improvement if one considers questions that were addressed directly and indirectly together (excluding Question 11) is 2.5 sigma, which is beginning to be interesting. There was no improvement on questions not addressed by the nighttime laboratories, which is assuring. Overall, this test shows evidence that the nighttime laboratories had an effect, but the statistics are too small to be definitive.

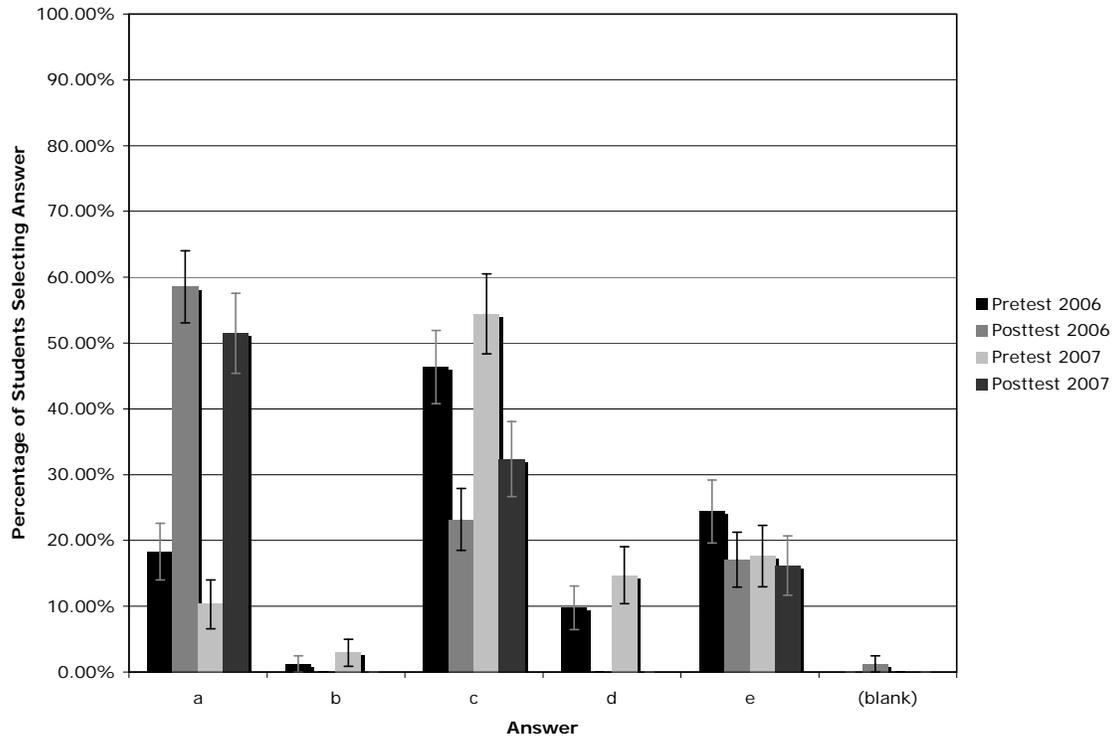

*Figure 2. Probability of obtaining a given answer for Question 4*

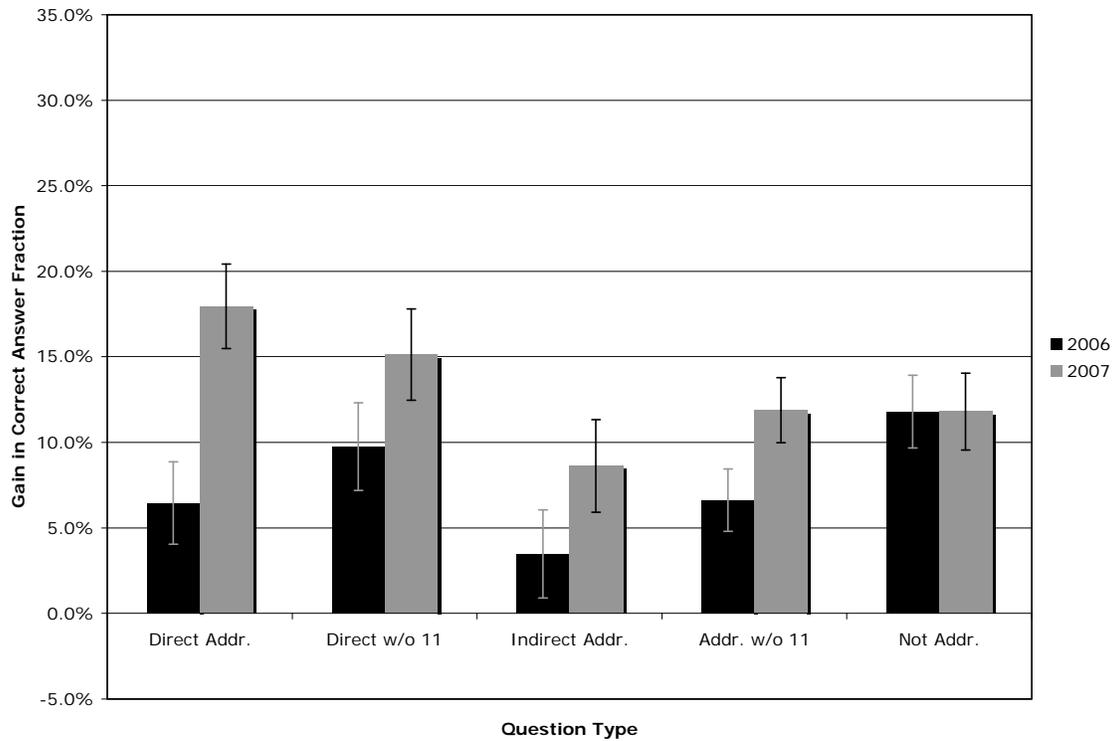

*Figure 3. Average change in the probability of obtaining the correct answer for questions addressed directly, addressed directly excluding Question 11, addressed indirectly, addressed directly or indirectly but excluding Question 11, and not addressed.*

| Q# | Lab Activity | | | | | | | | | | | | | | | | |
|---|---|---|---|---|---|---|---|---|---|---|---|---|---|---|---|---|---|
| | 1a | 1b | 1c | 2a | 2b | 2c | 3a | 3b | 3c | 4a | 4b | 4c | 5 | 6 | 7 | 8 | 9 |
| 1 | | | | | | D | | | | | | | | | | | |
| 2 | | | | | | | | | | | | | | | | | |
| 3 | | | | | | | | | | | | | D | | | | |
| 4 | | I | | | I | | | | | | | | | | | | |
| 5 | | | | | | | | | | | | | | | | | |
| 6 | | | | | | | | | | | | | D | | | | |
| 7 | | | | | | | | | | | | | I | | | | |
| 8 | | | | | | | | | | | | | | | | | |
| 9 | | | I | | | | | | | | | | | D | | | |
| 10 | | | | | | | | | | | | | I | | | | |
| 11 | | | | | | | | | | | | | D | | | | |
| 12 | | | | | | | | | | | | | D | | | | |
| 13 | | | | | | | | | | | | | | | | I | |
| 14 | | | | | | | | | | | | | | | | | |
| 15 | | | | | | | | | | | | | I | | | | |
| 16 | | | D | | | | | | | | | | | D | | | |
| 17 | | | | | | | | | | | | | | | | | |
| 18 | | | | | | | | | | | | | | | | | |
| 19 | | | | | | | | | | | | | | | | | |
| 20 | | | | | | | | | | | | | D | | | | |
| 21 | | | | | | | | | | | | I | | | | | |
| 22 | | | | | | | | | | | | | I | | | | |
| 23 | | | | | | | | | | | | | | | | | |
| 24 | | | | | | | | | | | | | I | | | | |
| 25 | | | | | | | | | | | | | | | | | |

*Table 3. Correspondence of assessment questions to lab activities covering them directly (D) or indirectly (I).*

Following the recommendation of Brogt et al. (2007), we will use multiple statistical techniques to better test the validity of our results. Because this analysis assumes that the students in the 2006 and 2007 classes are drawn from the same populations, and is therefore subject to uncertainty from our somewhat ambiguous determination of this assumption, we decided to subject the data to a further test that used only the students in each year who answered incorrectly on the pretest. One might expect that the students who did not initially know the correct answer might be a more uniform population.

We performed a $\chi^2$ analysis to determine whether the probability of getting the right answer on the posttest was the same in 2006 and 2007, using only students who did not answer the question correctly on the pretest. The data for each question, and for each examination period, is shown in Table 4. In column 6 of Table 4 we tabulate the change in the probability of getting the correct answer on the posttest, for students who did not get the question right on the pretest, between 2006 and 2007. For example, on Question 1, the fraction of students in this group who answered correctly on the posttest in 2006 was 18/(15+18)=0.545. In 2007, the fraction was 14/(10_14)=0.583, so the change in the probability of obtaining a correct answer was 3.8%. The questions with positive delta in this column are questions that showed improvement, while negative deltas indicate the 2007 results were worse. There was improvement in learning in 11 of the 15 questions addressed by the night labs; on the 10 questions that were not addressed, there were an equal number with higher and lower scores in 2007 compared to 2006.

The last column tabulates $\chi^2$, which is a measure of the probability that the years 2006 and 2007 have the same distribution. Since the number of exams in 2006 and 2007 are fixed, if the probability of getting the correct answer on the posttest is the same for both years then this is the only degree of freedom. The expected value of this probability is the total number of correct answers on the posttest divided by the total number of posttests. From this number we can calculate the expected number of 2006 posttests with incorrect answers, the expected number of 2006 posttests with correct answers, the expected number of 2007 posttests with incorrect answers, and the expected number of 2007 posttests with correct answers. $\chi^2$ is then the sum over these four numbers of $(N-<N>)^2/<N>$, where $N$ is the number and $<N>$ is the expected value of the number. If the value of $\chi^2$ is greater than 3.84, then the probability that such a difference would arise by chance is under 5%. The only question that shows a significant deviation is Question 11, with a $\chi^2$ value of 27.00, which arises by chance with probability only $2 \times 10^{-7}$. Performance on other questions also seems consistent with our previous results, but at lower significance as one would expect when data has been discarded to test for systematics.

Last, we explored potential differences between the four laboratory sections of the course that were held in 2007, which are labeled Tuesday 1 (T1), Tuesday 2 (T2), Wednesday 1 (W1), and Wednesday 2 (W2). The Tuesday sections were both taught by a team of one graduate and one undergraduate teaching assistant, and the Wednesday sections were taught by a different team of one graduate student and one undergraduate teaching assistant. Because the weather was different for these sections, they performed somewhat different laboratories. These four sections and the activities they performed are given in Table 5. Figures 4,5, 6, and 7 show the performance of each group separately.

Notice that the first Tuesday and the first Wednesday sections are fairly similar, while the second Tuesday section scored much higher and the second Wednesday section scored much lower. In fact, the second Wednesday section shows no improvement over the 2006 class. If the differences were between students on different days, we might have guessed that the teaching assistants differed in their effectiveness at delivering the material. But one does not see a significant difference between the two days. We might have assumed the second section for each might score better because the TAs were more prepared or worse because they were bored, but these results show that one day the second section had higher learning but on the other day the second section had a lower learning rate, so we would have to assume two such mechanisms to explain the results.

| Q# | Correct on neither 2006 test | Correct on 2006 Posttest only | Correct on neither 2007 test | Correct on 2007 Posttest only | Delta Probability Correct on Posttest only | $\chi^2$ |
|---|---|---|---|---|---|---|
| 1 | 15 | 18 | 10 | 14 | 3.8% | 0.08 |
| 2 | 40 | 22 | 41 | 11 | -14.3% | 2.82 |
| 3 | 34 | 22 | 20 | 25 | 16.3% | 2.65 |
| 4 | 33 | 11 | 25 | 6 | -5.6% | 0.33 |
| 5 | 14 | 13 | 9 | 6 | -8.1% | 0.26 |
| 6 | 10 | 24 | 5 | 11 | -1.8% | 0.02 |
| 7 | 11 | 20 | 8 | 18 | 4.7% | 0.14 |
| 8 | 37 | 26 | 24 | 22 | 6.6% | 0.46 |
| 9 | 57 | 17 | 47 | 14 | 0.0% | 0.00 |
| 10 | 12 | 17 | 6 | 14 | 11.4% | 0.66 |
| 11 | 42 | 8 | 9 | 24 | 56.7% | 27.00 |
| 12 | 20 | 19 | 9 | 18 | 17.9% | 2.09 |
| 13 | 5 | 18 | 0 | 6 | 21.7% | 1.58 |
| 14 | 5 | 9 | 1 | 6 | 21.4% | 1.05 |
| 15 | 46 | 16 | 33 | 16 | 6.8% | 0.63 |
| 16 | 35 | 12 | 27 | 21 | 18.2% | 3.48 |
| 17 | 1 | 16 | 1 | 9 | -4.1% | 0.16 |
| 18 | 25 | 23 | 18 | 26 | 11.2% | 1.15 |
| 19 | 10 | 16 | 8 | 11 | -3.6% | 0.06 |
| 20 | 2 | 3 | 4 | 1 | -40.0% | 1.67 |
| 21 | 29 | 11 | 15 | 12 | 16.9% | 2.05 |
| 22 | 16 | 25 | 11 | 21 | 4.6% | 0.17 |
| 23 | 45 | 5 | 53 | 6 | 0.2% | 0.00 |
| 24 | 40 | 12 | 32 | 14 | 7.4% | 0.68 |
| 25 | 25 | 29 | 26 | 18 | -12.8% | 1.59 |
| Direct | 215 | 123 | 131 | 128 | 13.0% | 10.22 |
| " w/o 11 | 173 | 115 | 122 | 104 | 6.1% | 1.92 |
| Indirect | 163 | 119 | 115 | 95 | 3.0% | 0.45 |
| Both w/o 11 | 336 | 234 | 237 | 199 | 4.6% | 2.12 |
| Unaddressed | 231 | 170 | 196 | 127 | -3.1% | 0.70 |

*Table 4. Fractional Change in Probability of Obtaining Correct Answers on the Posttest if the student's Pretest is had an Incorrect Answer.*

After making these plots, what we realized is that the second Tuesday section had all good weather, and the second Wednesday section had no clear nights of observing at all. We did not expect the weather to have a significant effect on learning, because practically all of the questions that were addressed by the laboratories were addressed in activities that did not require clear skies. Additionally, the students all did the parts of the labs intended for clear nights that addressed the questions on the assessment test. From a teaching point of view, we believe that all of the sections were exposed to all of the material required to do well on the assessment posttest.

We performed a statistical test to determine whether each section was different from the other three. Considering all of the categories in Figures 4-7, plus a fifth category that includes all 25 questions, we calculated the confidence level that the 2007 pretest results and the 2007 posttest results for each category in each section were different from the sum of the other three sections. For example, in the 2007 posttest, 10 of 17 students in section W2 got the correct answer for Question 11, and 43 of 48 students in the other three sections (combined) got the correct answer for Question 11. Assuming W2 was the same as all other sections, we would have expected 13.9 students to get the correct answer and 3.1 students to get the wrong answer. In the other three sections we expect 39.1 students to get the correct answer and 8.9 students to get the wrong answer. $\chi^2$ is calculated as $(10-13.9)^2/13.9+(7-3.1)^2/3.1+(43-39.1)^2/39.1+(5-8.9)^2/8.9=7.89$. Such a $\chi^2$ value arises by chance with probability only 0.5% (confidence level of 100%-0.05% = 99.5%), thus the conclusion that the W2 section scored differently from the other three sections on this question is statistically significant. The resulting confidence levels that each exam and category in each section is different from the same exam and category for the other three sections is shown in Table 6. The only confidence levels over 90% are in the posttests for the T2 and W2 sections. Red numbers indicate that the section scored significantly worse than the other sections and green numbers indicate that the section scored significantly better. There were no unusual differences between the pretests. All cases of a section doing significantly better were in T2, and all cases of a section doing significantly worse are in W2.

These results give us a strong suspicion that the sections were not the same, and we rationalize this by pointing to the weather as a major factor in the outcomes between sections, since that is the only single factor that we could find that explained all of the results. It should be stressed, though, that our experiment was not designed to test this hypothesis, and the result is therefore the result of multiple hypothesis testing. Further studies should be designed to test the relationship between clear weather and learning.

| Section | Lab Activity | | | | | | | | | | | | | | | | |
|---|---|---|---|---|---|---|---|---|---|---|---|---|---|---|---|---|---|
| | 1a | 1b | 1c | 2a | 2b | 2c | 3a | 3b | 3c | 4a | 4b | 4c | 5 | 6 | 7 | 8 | 9 |
| Tuesday 1 | X | X | X | X | X | X | | X | X | | | | X | | X | X | |
| Tuesday 2 | X | X | X | X | X | X | X | X | X | | | | X | | X | X | |
| Wednesday 1 | X | X | X | | X | X | X | X | X | | | | X | X | X | | X |
| Wednesday 2 | * | * | * | * | * | * | | | | | | | X | X | X | X | X |

*Table 5. Correspondence of laboratory sections to activities performed. \* represents activities performed that were not actually done in full due to cloudy skies.*

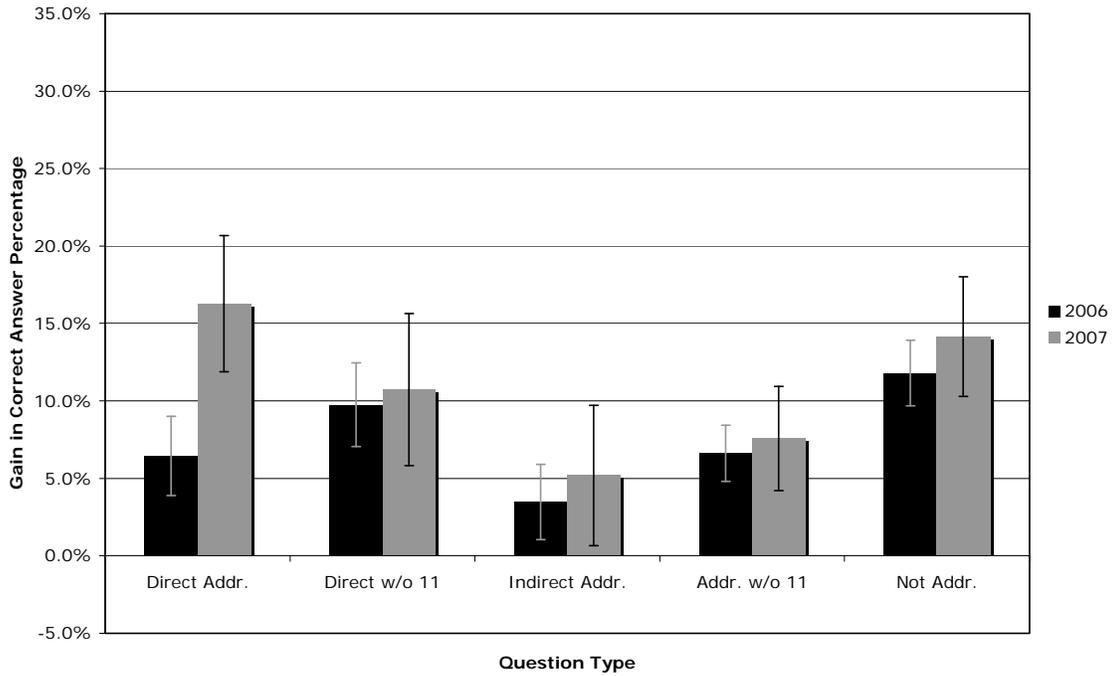

*Figure 4. Average change in the probability of obtaining the correct answer for questions addressed directly, indirectly, and not addressed for the Tuesday 1 section.*

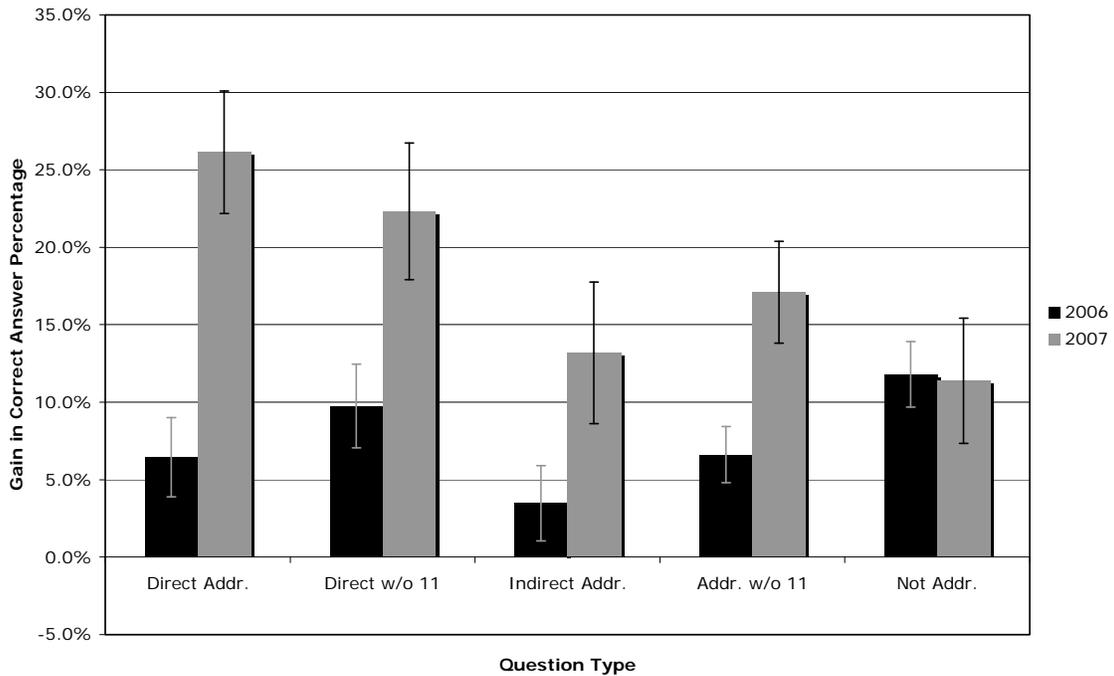

*Figure 5. Average change in the probability of obtaining the correct answer for questions addressed directly, indirectly, and not addressed for the Tuesday 2 section.*

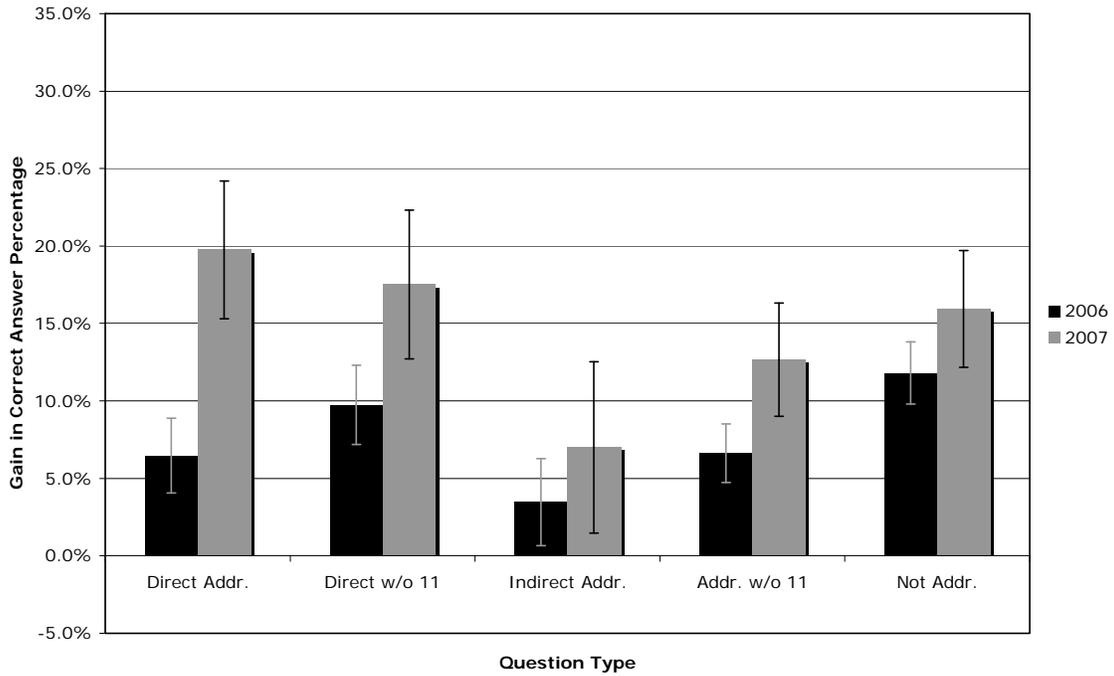

*Figure 6. Average change in the probability of obtaining the correct answer for questions addressed directly, indirectly, and not addressed for the Wednesday 1 section.*

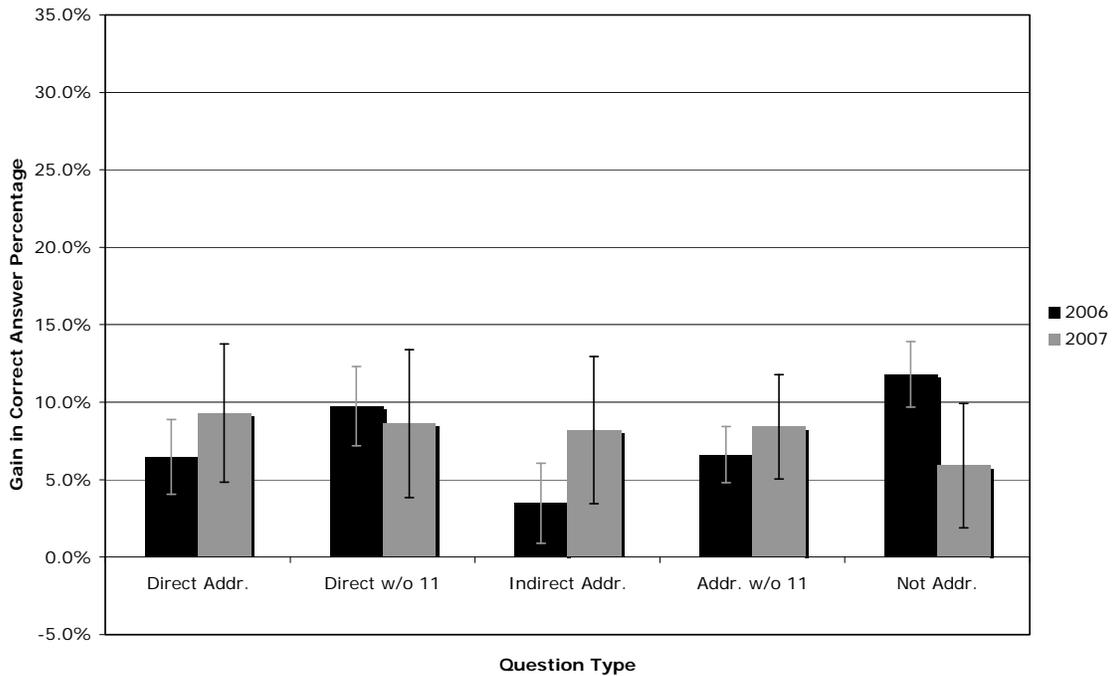

*Figure 7. Average change in the probability of obtaining the correct answer for questions addressed directly, indirectly, and not addressed for the Wednesday 2 section.*

One fact that caused us to question the validity of this test is that the pretest results for Question 11 in section W2 were low compared to the other sections. In fact, if the posttest results for each section are compared to the pretest results for the same section (instead of averaging the pretest results under the assumption that all sections start out the same), we found that the learning on Question 11 was lowest for the W1 section. Since the difference in posttest results for this question are the most dramatic, and a different analysis of the data throws doubt on the result, this is further cause to require further testing of the hypothesis that weather affects learning in nighttime laboratories. We note, however, that the overall pretest scores are not unusual for section W2, so we cannot conclude which analysis is more correct.

| Q# | Pre/Post Test | Tuesday Section 1 | Tuesday Section 2 | Wednesday Section 1 | Wednesday Section 2 |
|---|---|---|---|---|---|
| 11 | pretest | 56% | 48% | 84% | 86% |
|  | posttest | 88% | 85% | 14% | **99.5%** |
| Direct | pretest | 14% | 32% | 61% | 84% |
|  | posttest | 52% | **99.2%** | 43% | **98.4%** |
| " w/o 11 | pretest | 8% | 51% | 31% | 66% |
|  | posttest | 78% | **98.0%** | 48% | **90.7%** |
| Indirect | pretest | 32% | 62% | 32% | 60% |
|  | posttest | 58% | 43% | 15% | 5% |
| Both w/o 11 | pretest | 28% | 73% | 43% | 79% |
|  | posttest | 85% | **95.8%** | 45% | 74% |
| Unaddressed | pretest | 30% | 70% | 21% | 30% |
|  | posttest | 57% | 6% | 75% | **92.9%** |
| All Questions | pretest | 10% | 3% | 62% | 68% |
|  | posttest | 25% | **91.3%** | 75% | **98.6%** |

*Table 6. Confidence level of the detection of a difference between each section and the other three. Red numbers indicate the section scored significantly worse than the other three, and green numbers indicted the section scored significantly higher. The only comparisons that differ at more than 90% confidence are for Tuesday Section 2, which in all cases did better than expected; and for Wednesday Section 2, which in all cases did more poorly than expected.*

4.2. Solicited Feedback and Course Evaluations

An informal mid-term survey, given by the professor of the lecture portion of the course in 2007, received 34 replies of a possible 95 students register for the class at the time. Raw scores were largely neutral, with a rating on the labs as a learning experience obtaining 3.4/5.0, interest generated by labs as 3.3/5.0, organization of labs obtaining 3.2/5.0, and a rating of the TAs' helpfulness at 3.8/5.0.

Twenty-two written comments were received with the surveys. Of these, two comments complained about the lack of good weather, two comments were generally positive, eight were neutral (including two that expressed concerns about the labs being 'rushed' due to other students' desires to not remain in the labs for an extended period) and eight were negative.

In addition to this survey that attempted to assess the usefulness of the labs themselves, the long form of the IDEA course assessment form is regularly given to students at the end of each course at Rensselaer. Unfortunately, in Fall 2006 only 37 out of 88 registered students filled out the IDEA forms. While the Fall 2007 IDEA form results are "highly reliable," as 70 out of 88 students responded, it is difficult to compare them with the results from 2006, before the laboratories were implemented.

Nonetheless, we note that the "Excellent Teacher" rating remained unchanged between the two semesters; it was 4.0/5.0 both times. When the laboratories were implemented, the "Progress on Relevant Objectives" rating increased slightly from 3.1/5.0 to 3.4/5.0 and the "Excellent Course" rating decreased from 3.8/4.0 to 3.3/5.0. In all cases, we are quoting the IDEA "adjusted score," but the raw scores are not very different.

In 2006, only eight students wrote comments, all of which were extremely positive about the teacher, the class, the material, and the laboratories. In 2007, fourteen students wrote comments (about the same percentage). The comments on the quality of the course and instructor were about evenly split, and of the eight that mentioned the laboratories, three complained that they did not get any good weather, one comment was positive, three were negative, and one was neutral. From these results, we feel that there was no large change in opinions caused by the introduction of the laboratory activities; if anything, there was a decrease in the students' rating of the course as a result of their introduction even though, by the students' own estimation, the progress on the course objectives was, if anything, higher.

5. DISCUSSION AND CONCLUSIONS

We tested the effectiveness of night-time laboratories (http://www.rpi.edu/dept/phys/observatory/labs.html) on learning in a non-major introductory astronomy class at Rensselaer Polytechnic Institute, using a 25-question assessment test developed for this study. Of the 25 questions, eight were directly addressed by the laboratories, seven were indirectly addressed by the laboratories, and ten were not addressed in the laboratory, but were addressed in the lecture portion of the class, which remained unchanged between 2006 and 2007. We suggest that the fourth question on our assessment test should be discarded or re-worded, as it generates ambiguous results that are hard to interpret.

For one of the directly addressed questions, a geometrical question which tested whether students know that the full moon cannot be seen at noon, there was an enormous (eight sigma) improvement in learning after the introduction of the nighttime laboratories. On the pretest, 45% of the combined 2006 and 2007 classes chose the correct answer. In 2006, only 28% of the class had the right answer on the post test, but in 2007, 82% of the class was able to answer this question correctly. We attribute this improvement to the cloudy activity "Light and Shadow in the Solar System," which was introduced to the 2007 class. This result is obtained even if we only include the posttests of the students who answered the question incorrectly on the pretest, and is a result with extreme statistical significance.

There was general improvement in the learning of the other fourteen questions addressed by the nighttime laboratories, but at low significance. There was no overall change in learning on questions that were not addressed by the night laboratories; simply introducing new laboratories did not stimulate the students to study the general course material at a higher level. There are some indications that the course was more effective in meeting the learning objectives even from the students' point of view, but there was an overall negative reaction to being asked to think in night laboratories, as evidenced by a small reduction in the students' perception of "excellence of course" on the IDEA forms, and particularly for the lab section that did not have a single clear night all semester. Although the statistics are not overwhelming, there is evidence that the student populations in both years and all sections of the class were similar.

It is interesting to note that the one section out of four in 2007 that showed slightly less learning in all categories was the section that never had a clear night. Additionally, the section with the best weather showed the highest learning achievement. These results are suggestive, but are subject to a critique of multiple hypothesis testing; furthermore, the results change somewhat with different analysis techniques, so they should be specifically tested for in future learning studies. If we assume our result is correct, we conclude that the atmosphere in which the material is presented has a strong effect on whether the students learn the material. All of the sections had nearly the same exposure to the portions of the activities that addressed the learning objectives on the assessment test. The questionnaires distributed to the students and the testimony of the graduate teaching assistants indicate to us that the students in the section that got no clear weather were discouraged by their experience, and there is some evidence that this translates into poorer retention of the course material.

We pioneered the idea of laboratory rotations in the nighttime laboratories, and encountered significant obstacles in the required night time staffing level, the amount of time required from the students taking the course, timing the switch between stations, and coordinating the chaos of running a different laboratory every night. We found that undergraduate and first-year graduate student teaching assistants found running the laboratories extremely challenging. The students were not prepared to be out until 10:00 or 11:00 PM working in the observatory on their assigned nights. It was difficult to time all three activities so that they took the same amount of time; any equipment difficulties on the 16" telescope put that portion of the laboratory way behind schedule. And, since each section had a different number of clear nights, and we needed both a clear and a cloudy laboratory ready for any given night, it was challenging to prepare to teach these classes. Hemenway et al. (2002), found that results may vary between the first and second semesters that a class has implemented activities-based learning, with the second semester of activity-based learning showing more significant improvement in both astronomy content and the course survey. Given the administrative difficulties we encountered when introducing this new curriculum, we would not be surprised to find that learning and course survey results improve the second time around for this set of activities as well. It is clear that more effort should be placed in training teaching assistants who run the night laboratories.

In the future, we plan to reduce the stress of the night time laboratories on both the teaching assistants and the students, while keeping those activities that have been shown to significantly improve student learning. Instead of using rotations through three stations, we plan to schedule smaller groups to be in the observatory for shorter periods of time. During the time they are in the observatory, students will be working with a teaching assistant in a smaller group, so that they will spend less time waiting. It turned out to be unrealistic to expect non-majors to quickly learn how to operate the small telescopes, so laboratories that required this skill will be modified or eliminated. With shorter laboratories, which the students attend every week rather than every other week, there will be less variation in the amount of clear weather that each section experiences.

We expended considerable effort and expense to refurbish our campus observatory so that could be used for educational and outreach purposes, and to generate new laboratories that made it possible for many more students to look through real telescopes, and to see how modern astronomers view the sky with CCD cameras. This study shows that we were able to obtain significant improvement with hands-on activities that did not require telescopes, and we showed weak evidence that students who were able to look through the telescope had a better attitude towards the class, which resulted in higher learning rates. Additionally, we now have an observatory that supports our advanced undergraduate students in learning to be astronomers, will allow people with disabilities to become involved with public observing and coursework, and which has re-energized the student astronomy club; the club members have run their own events, help with weekly public observing events, train new members to use the telescopes, and have maintained the electro-mechanical and computer components of the observatory. Overall, the project has had a significant impact on the campus and community.

This work was supported by NSF grant DUE 05-11340 and the NASA/NY Space Grant.

# Astronomy Survey

## Circle the best answer to each of the questions below.

1. Compared to the Earth's radius (4000 miles), the thickness of the Earth's atmosphere is
    a. greater than 4000 miles.
    b. less than 4000 miles.
    c. approximately 4000 miles.

2. The spectrum of a star can provide information about all of the following *except*
    a. the star's temperature.
    b. the star's rotation rate.
    c. the star's chemical composition.
    d. the star's speed along the observer's line-of-sight.
    e. the star's speed perpendicular to the observer's line-of-sight.

3. The same side of the Moon always faces the Earth primarily because
    a. the Moon does not rotate on its axis.
    b. the Moon does not orbit the Earth.
    c. the Moon's rotational and orbital periods are equal.
    d. the Earth and the Moon have equal rotational periods.
    e. the Moon is in a geo-synchronous orbit around the Earth.

4. Ignoring the effect of the Earth's atmosphere, the *resolving power* (ability to distinguish objects that are close together) of a telescope is dependent upon
    a. the size of the aperture accepting the light only.
    b. the color of the light being observed only.
    c. the size of the aperture accepting the light and the color of the light being observed.
    d. the color of the light being observed and the size of the light source.
    e. the length of the telescope and the size of the light source.

5. It is necessary to add an extra day to the calendar every 4 years (leap year) primarily because
    a. one year is actually a little more than 365 days.
    b. February, on average has too few days.
    c. one year is actually a little less than 365 days.
    d. it is the only way for the solar calendar to remain in sync with the lunar calendar.
    e. the Earth's rotation rate slows by about a day during this period of time due to tidal interactions with the Moon.

6. The seasonal changes that occur in New York State are caused primarily by the fact that the
    a. Earth is closer to the Sun in the summer and further from the Sun in the winter.
    b. Earth is closer to the Sun in the winter and further from the Sun in the summer.
    c. Earth's equatorial radius is slightly larger than its polar radius.
    d. Earth's rotational axis is tilted with respect to the Earth's orbit around the Sun.
    e. Earth rotates on its axis.

7. An observer views a star at a certain distance and notes its brightness. If the observer were to increase his/her distance from the star, the star's *observed* brightness would
    a. increase.
    b. decrease.
    c. remain the same.

8. Over a six-month period an observer on the Earth records the position of a nearby star. The observer notes that at the end of the six-month period the star has slightly shifted its position in the sky. This observed shift in the star's position could allow the observer to determine the Earth's
    a. mass.
    b. rotation rate.
    c. radius.
    d. orbital tilt.
    e. distance from the star.

9. What property of the North Star allows it to be used for navigation in the northern hemisphere?
    a. It is the brightest star in the sky, which makes it easy to find at night.
    b. It is located at the same altitude in the sky everywhere in the northern hemisphere.
    c. Its altitude is approximately equal to an observer's latitude.
    d. Very bright stars surround it, which makes it easy to locate.
    e. Both a and b.

10. Assume a new Moon occurred on January 1st. Which diagrams below correctly show the phase of the Moon on the evenings of January 4th and January 8th?

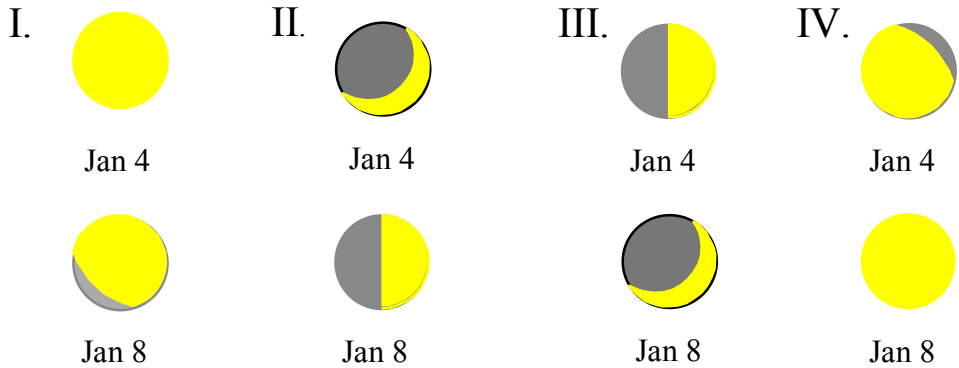

(Note: The lighter shaded side is the illuminated side in the diagrams.)

   a. I.
   b. II.
   c. III.
   d. IV.

11. A full Moon can never be seen at
   a. sunrise.
   b. sunset.
   c. noon.
   d. midnight.
   e. It is possible to see the full Moon at any time of the day.

12. In New York State, eclipses of the Sun are not visible every month due to the fact that
   a. the Moon is not visible in New York State every month
   b. the Moon is rarely in the sky during the daytime.
   c. the Moon is generally too far from the Earth to block out the Sun.
   d. the Moon's orbit is tilted with respect to the Earth's orbit around the Sun.
   e. cloud cover generally obscures our view.

13. Given that the color of a star is directly related to the star's *surface temperature*, it would follow that, compared to a star that is red in color, a star that is blue in color is
   a. at a higher surface temperature than the red star.
   b. at a lower surface temperature than the red star.
   c. at the same surface temperature as that of the red star.

14. The Moon's influence on the Earth's ocean tides is more significant than that of the Sun's primarily because
    a. the Moon is so much larger than the Sun in the sky.
    b. the Moon is so much closer than the Sun.
    c. the Moon can diminish the affect of the Sun's gravitational field during the new and full moon phases.
    d. the Moon's distance from the Earth, unlike the Sun's remains roughly constant.
    e. The statement is actually false, as the Sun being significantly larger clearly affects the ocean tides more than the Moon.

15. An observer in New York State records the compass directions of the rising and setting Sun. During the period from March 20 (first day of spring) to June 20 (first day of summer) the observer notices that each day the Sun
    a. rises more toward the northeast and sets more toward the northwest.
    b. rises more toward the northeast and sets more toward the southwest.
    c. rises more toward the southeast and sets more toward the northwest.
    d. rises more toward the southeast and sets more toward the southwest.
    e. always rises due east and sets due west.

16. In New York State we observe different constellations in the evening sky at different times of the year. This is primarily due to the fact that the
    a. Earth's axis is tilted with respect to the Earth's orbit around the Sun.
    b. Earth rotates on its axis.
    c. Earth wobbles on its axis much like a spinning top.
    d. Earth revolves around the Sun.
    e. Earth's magnetic field prevents the viewing of certain constellations throughout the year.

17. The *intrinsic* brightness of a star depends on its surface temperature and its size. The equation that governs this relationship is given by $L = kR^2T^4$, where L is the intrinsic brightness of the star, R is the radius of the star, T is the star's surface temperature and k is a constant. Given this relationship, an extremely bright star that is relatively cool must be
    a. relatively large.
    b. relatively small.
    c. about the size of the Sun.
    d. a black hole.
    e. a white dwarf.

18. A nearby star is moving directly toward the Earth.  Compared to the spectral lines of the star if it were not in motion, the spectral lines of the star due to its motion toward the Earth are
    a. shifted toward the blue end of the spectrum because the star is getting brighter as it approaches the Earth.
    b. shifted toward the red end of the spectrum because the star is getting brighter as it approaches the Earth.
    c. shifted toward the blue end of the spectrum because the light waves get compressed as the star approaches the Earth.
    d. shifted toward the red end of the spectrum because the light is perceived to move faster as the star approaches the Earth.
    e. unchanged as the spectral lines of a star are not affected by the star's motion toward or away from an observer.

19. Jupiter is 5 times as far from the Sun as the Earth.  Compared to the time it takes the Earth to orbit the Sun (1 year), the time it takes Jupiter to orbit the Sun is
    a. 5 years.
    b. less than 5 years.
    c. more than 5 years.

20. The illumination of the Moon's surface is primarily caused by
    a. light from distant stars.
    b. light from the Sun.
    c. light from the Earth.
    d. the Moon's ability to produce its own light.
    e. the culmination of light from nearby planets.

21. The Earth moves at different speeds as it orbits the Sun.  This is caused primarily by
    a. the fact that the Earth's axis is tilted either toward or away from the Sun.
    b. the Moon's gravitational affect on the Earth in its orbit around the Sun.
    c. a frictional force due to the interaction of the Earth's magnetic and gravitational fields.
    d. the fact that the Earth is sometimes closer to the Sun in its orbit.
    e. The statement is actually false, as the Earth maintains a constant speed as it orbits the Sun.

22. Compared to the maximum altitude of the Sun at the Tropic of Cancer (latitude = $23.5^0$ north) on June 20, the maximum altitude of the Sun in Albany, New York (latitude = $42^0$ north) is
    a. higher.
    b. lower.
    c. the same.

23. It is observed that a vertical tree in city A casts no shadow at noon on June 20. However, a vertical tree in city B that is due South of city A does cast a shadow at noon on June 20. If the angle of the shadow and the distance between city A and B are known, this information could lead to a measurement of the Earth's
    a. orbital tilt.
    b. mass.
    c. distance from the Sun.
    d. circumference.
    e. rotation rate.

24. An observer on the Earth views a total eclipse of the Moon. An observer on the Moon who is simultaneously viewing the Earth would see a
    a. total eclipse of the Earth only.
    b. total eclipse of the Sun only.
    c. partial eclipse of the Earth only.
    d. partial eclipse of the Sun only.
    e. total eclipse of both the Sun and the Earth.

25. According to Hubble's Law, galaxies that are twice as far from the Earth are on average moving twice as fast away from the Earth. The equation that governs this motion is given by v = Hd, where v is the average speed of all the galaxies at a given distance, d is the distance from the Earth and H is a constant. As a consequence of this relationship astronomers have been able to estimate the
    a. critical density of the universe.
    b. mass of the universe.
    c. number of galaxies in the universe.
    d. age of the universe.
    e. Earth's location in the universe.

LITERATURE CITED

Adams, J. P., & Slater, T. F. 1998, "Using Action Research To Bring the Large Lecture Course Down to Size," *Journal of College Science Teaching*, **28**(2), 87-90.

Allen, M. L. and Kelly-Riley 2006, "Promoting Undergraduate Critical Thinking in Astro 101 Lab Exercises," *Astronomy Education Review*, **4**(2), 10-19.

Brogt, E., Sabers, D., Prather, E. E., Deming, G. L., Hufnagel, B., and Slater, T. F. 2007 "Analysis of the Astronomy Diagnostic Test," *Astronomy Education Review*, **6**(1), 25-42.

Gould, R., Dussault, M., and Sadler, P. 2007, "What's Educational about Online Telescopes?: Evaluating 10 Years of MicroObservatory," *Astronomy Education Review*, **5**(2), 127-145.

Hemenway, M. K., Straits, W. J., Wilke, R. R., & Hufnagel B. 2002, "Educational Research in an Introductory Astronomy Course," *Innovative Higher Education*, **26**(4), 271-280.

Hoyt, D. P., & Lee, E. 2002, "Technical Report No. 12: Basic Data for the Revised IDEA System," http://www.idea.ksu.edu/reports/techreport-12.pdf.

McCrady, N. and Rice, E. 2008, "Development and Implementation of a Lab Course for Introductory Astronomy," *Astronomy Education Review*, **7**(1).

Slater, T. 2008, "The First Big Wave of Astronomy Education Research Dissertations and Some Directions for Future Research Efforts," *Astronomy Education Review*, **7**(1).

Straits, W. J., & Wilke, R. R. 2003, "Activities-based Astronomy: An Evaluation of an Instructor's First Attempt and its Impact on Student Characteristics," *Astronomy Education Review*, **2**(1), 46-64.

Zeilik, M. 2003, "Birth of the Astronomy Diagnostic Test: Prototest Evolution," *Astronomy Education Review*, **1**(2), 46-52.